\documentclass[12ptxs]{article}
\usepackage{blindtext}
\usepackage[T1]{fontenc}
\usepackage[utf8]{inputenc}
\DeclareUnicodeCharacter{2212}{-}
\usepackage{amsmath}
\usepackage{graphicx}
\graphicspath{{images/}}
\usepackage{subfigure}
\usepackage{hyperref}
\usepackage{authblk}
\usepackage[normalem]{ulem}
\usepackage[stable]{footmisc}
\usepackage[superscript]{cite}
\usepackage{comment}
\hypersetup{
    colorlinks=true,
    linkcolor=blue,
    filecolor=magenta,      
    urlcolor=cyan,
}
\usepackage{vmargin}
\setmarginsrb{2.5 cm}{1.5 cm}{2.5 cm}{1.5 cm}{0.5 cm}{0.5 cm}{0.5 cm}{0.5 cm}

\date{\today}
\usepackage{setspace}
\usepackage{color,soul}

\title{\bf{Microphase Separation Controls the Dynamics of Associative Vitrimers}}
\author[1,2]{Rahul Karmakar}

\author[1,2]{Abhishek S. Chankapure}

\affil[1]{Department of Chemical Engineering, Indian Institute of Technology Madras, Chennai 600036, India\\
}

\affil[2] {Center for Atomistic Modeling and Materials Design, Indian Institute of Technology Madras, Chennai 600036, India \\
}

\author[3]{Srikanth Sastry}
\affil[3]{Theoretical Sciences Unit and School of Advanced Materials, Jawaharlal Nehru Centre for Advanced Scientific Research, Rachenahalli Lake Road, Bengaluru-560064, India\\
}

\author[4]{Sanat K. Kumar}

\affil[4]{Department of Chemical Engineering, Columbia University, New York, USA\\
}%
\author[1,2]{Tarak K. Patra}

\begin{document}

\maketitle

\begin{abstract}
\noindent
Vitrimers are a class of polymers characterized by dynamic covalent networks, where specific monomer units— ``stickers" —form reversible crosslinks that enable network rearrangement without loss of overall connectivity. The conventional wisdom is that the sticker dynamics control chain relaxation behavior, and hence the mechanical properties of associative vitrimers where the number of crosslinked sticker pairs is precisely constant over time. Instead, here we show that the chemical differences between sticker groups and non-sticky chain monomers can cause them to microphase separate. Under these conditions, the slow exchange of a sticker from one microphase to an adjacent one controls relaxation behavior. Controlling sticker aggregation is thus a key to tailoring the properties of these polymers with immediate relevance to a circular polymer economy.
\end{abstract}

\newpage

\doublespacing

While permanently crosslinked polymers are ubiquitous due to their favorable mechanical properties, they cannot be reprocessed. In contrast, polymer networks that can rearrange their topology without depolymerization, e.g., via the formation of reversible crosslinks or by having exchangeable crosslinks, are readily reprocessable especially at high temperature, while maintaining their favorable properties under normal use conditions. Understanding the behavior of these constructs is hence an important research topic \cite{montarnal2011silica,lu2012making,jin2013recent,samanta2021polymers,maaz2021synthesis,long2013modeling,meng2016stress,snyder2018reprocessable,semenov1998thermoreversible,rubinstein1998thermoreversible,porath2022vitrimers,zou2018rehealable,lewis2023effects}. While the dominant role of the crosslink moieties is to provide the network-like connectivity to these materials, another salient, less appreciated feature, is that these groups are chemically quite dissimilar to the chain monomers. These chains with groups having significantly different polarizabilities can thus exhibit microphase separation leading to the formation of  hierarchical nano- and microstructures\cite{zhang20171,ricarte2018phase,ricarte2020linear, arbe2023microscopic}. 
We posit that this combination of reversible crosslinking and microphase separation leads to strongly enhanced mechanical behavior relative to the base, unfunctionalized polymer. To this point, Ricarte et al. \cite{ricarte2018phase} experimentally showed the formation of self-assembled, percolating nanostructures within the graft-rich regions —
these form a continuous network which supports the material’s solid-like behavior and its dynamic exchange-based flow under applied stress. Complementing this, Zhang and Waymouth \cite{zhang20171} reported on the self-assembly and reversible crosslinking of dithiolane-functionalized block copolymers in aqueous systems. Their study highlights how the balance between dynamic covalent chemistry and micelle self-assembly dictates both the network topology and mechanical adaptability of the resulting hydrogels. These studies indicate that the lifetimes of the dynamic crosslinks and microphase separation together govern the relaxation behavior of polymer chains, which is directly linked to the mechanical and dynamic properties of these systems. Earlier  simulation studies have reported the role of dynamic crosslinks on the chain relaxation and mechanical properties wherein the crosslinks are well mixed in polymer melts.\cite{zhao2022molecular,zhao2023unveiling,perego2022microscopic, perego2021effect, xia2024effect, wu2019dynamics,amin2016dynamics,nie2023competing} However, the molecular mechanism of the chain relaxations in experimentally relevant, microphase separated, reversible polymer networks is not well-studied.

With these implications in mind here we study the chain relaxation dynamics of a coarse-grained model polymer melt with exchangeable crosslinks using a hybrid molecular simulation method. The polymer chains are composed of non-reactive and reactive (stickers) beads, with the stickers capable of forming bonds with other stickers reversibly, with bond breaking occurring through an exchange of bonding partners. The bond exchange is simulated using a Monte Carlo (MC) scheme, while molecular dynamics (MD) simulations are conducted to relax the system. Specifically, we model an associative network, so the total number of reversible bonds remains constant as a function of time, and examine both their static and dynamic properties. From a static standpoint, the stickers distribute uniformly in the melt when the crosslinked monomers (stickers) and non-crosslinked monomers are compatible (i.e., all monomer pairs interact via the same Lennard Jones (LJ) potential). For the incompatible case, which is modeled by having dissimilar pairs interact {\it via.} the Weeks-Chandler-Andersen (WCA) potential, we observe segregation behavior controlled by the sticker distribution on a chain. If all monomers on a chain can engage in crosslinking interactions, while the number of crosslinks are fixed, then sticker positions on the chain are floating (``annealed"). In this ``annealed" sticker limit we find macrophase separation between the stickers and the monomers. In contrast,  when the sticker locations on the chains is fixed (``quenched" case), then we find microphase separation, with the specifics depending on the architecture of the chains (i.e., stickers evenly distributed on a chain, vs. a diblock copolymer wherein all the stickers are located at one of the ends of a chain, vs. a triblock coplymer with two sticker-blocks at the two ends of a chain). We characterize the cluster size and its relationship to chain relaxation behavior. In the case of the uniformly distributed stickers with same interactions across all monomers we find simple chain relaxation behavior, i.e., in the absence of sticker clustering. In contrast, the chain relaxation time shows a strong correlation with the cluster size for systems displaying microphase separation. While we expect that the dynamics of these materials are affected by the sticker lifetimes, these results underscore the importance of the compatibility of crosslink motifs and chain architecture in designing dynamic polymer networks with tunable structural and relaxation properties.

We employ the Kremer-Grest bead-spring polymer model\cite{kremer1990molecular} and perform hybrid MD – MC simulations in the isothermal-isobaric ensemble, as implemented in our previous studies \cite{karmakar2025computer,karmakar2025rouse}. In this model, two successive segments in a chain are connected by the finitely extensible nonlinear elastic (FENE) bond: $V_{bond}=-0.5kR_{0}^{2}\ln[1-(\frac{r}{R_{0}})^{2}]+4\epsilon[(\frac{\sigma}{r})^{12}-(\frac{\sigma}{r})^{6}]+\epsilon$ where the 2nd term in $V_{bond}$ is truncated at a cutoff distance $r_c = 2^{1/6}\sigma$, with $R_{0}=1.5\sigma$ and $k=30\epsilon/\sigma^{2}$. Here, $\sigma$ and $\epsilon$ are the size and cohesive interaction strength of a segment. All non-bonded monomers interact through a standard LJ potential truncated at 2.5$\sigma$. We consider 200 polymers each of chain length $N$=50. Since the entanglement chain length of this model polymer is $N_e \approx70$, we expect these chains to follow Rouse dynamics. %There are 200 chains in the box. %Therefore, the total number of particles in our simulations is 10000. 
There are two types of beads in a chain - non-reactive (``monomers") and reactive (``stickers"). The reactive beads participate in reversible crosslink formation.  %The sticker-sticker and monomer -monomer interaction is modeled by the same Lennard-Jones (LJ)  potential $V(r)=4\epsilon[(\frac{\sigma}{r})^{12}-(\frac{\sigma}{r})^{6}]$ with the cut-off distance of $2.5\sigma$. 
The cross-interaction between sticker-monomer is also modeled via the LJ potential when they are chemically compatible.  The incompatible crosslink-monomer interaction is represented as the WCA potential, which is the LJ potential with a cutoff distance of $2^{1/6}\sigma$. All of the interaction potentials are shifted to zero at their respective cutoff distances. 

We study two cases: 1) Randomly positioned stickers or the ``Annealed" case:   Any monomer bead can participate in a crosslink, but with the the total number of monomers participating in crosslinks being fixed at 5\% of the total beads in the system. When a bead forms a crosslink we change it's type from type ``monomer" (or type-1)  to ``sticker" (type-2) and vice-versa. 2) Fixed stickers (``quenched"): In each chain, we fix 3 beads as those that can participate in cross-linking.  So, 6\% of the beads are designated as stickers and among them 5\% are crosslinked. The stickers beads are always of type-2 here throughout the simulation.
We study three such architectures: a) {\em Evenly spaced}: Here 3 stickers are equally spaced on the chain, i.e., they are the 9$^{th}$, 25$^{th}$, and 41$^{th}$ beads of a chain. b) {\em Triblock}: Two stickers are as beads 1 and 2, and a third one is in the $50^{th}$ position at the end of a chain. c) {\em Diblock}: The stickers are on beads 1,2 and 3 on each chain.

\begin{figure}[!htb]
    \centering
     \includegraphics[width=16.3cm,height=10.3cm]{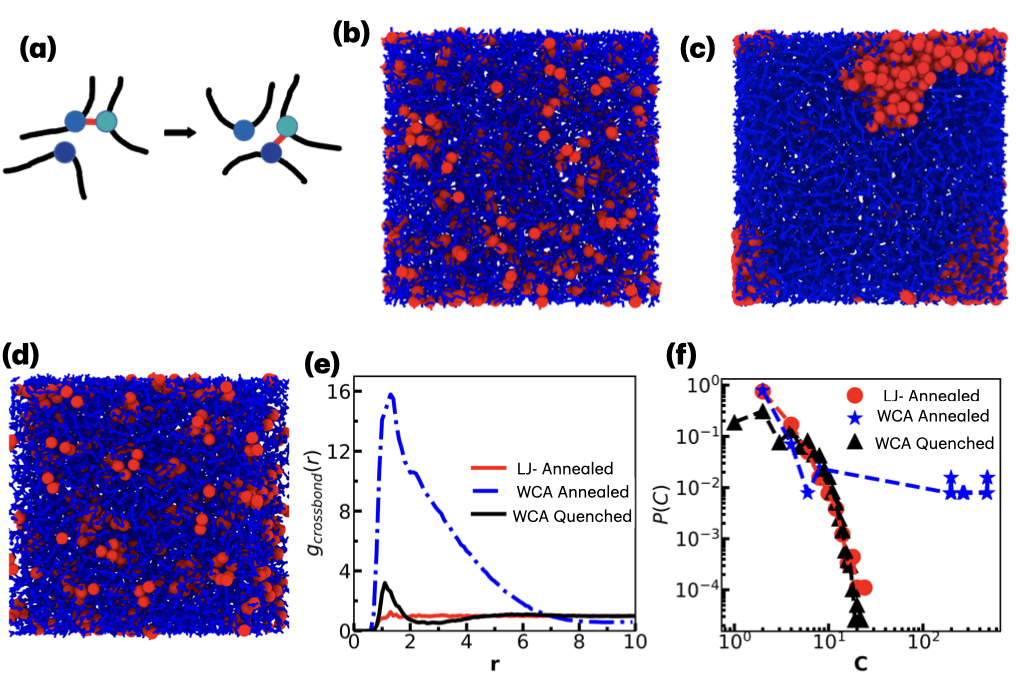}
    \caption{(a) A schematic representation of bond exchange mechanism. (b) An MD snapshot of the system with the same LJ interaction between all moieties. (c) MD snapshots of the system with WCA interaction between sticker-monomer for the case of annealed stickers. (d) Corresponding case with quenched stickers with evenly spaced stickers along a polymer chain. (e) The pair correlation function between stickers, and (f) the corresponding cluster size distribution. All the data correspond to $T$=1.0.}
    \label{graph1}
\end{figure}

 We set up our simulation box with periodic boundary conditions in all three directions. MD simulations are conducted using the LAMMPS open-source code.\cite{LAMMPS}. The MD time between two consecutive MC cycles is defined as $\tau_{c}=5\tau$ (i.e., 1000 MD steps). Here, $\tau=\sqrt{(m\sigma^{2}/\epsilon)}$ is the unit of time, $m$ being the mass of a bead. The $\sigma$ and $\epsilon$ are the size and cohesive energy of the LJ interaction, respectively. During each MC cycle, we perform 200 MC moves. %Here we use associative network: so total bonds in the systems are constant. 
 To exchange bonds, first we choose an existing crosslink bond between two stickers, labeled as $i$-$j$. We then randomly select one of the stickers in this bond, say $i$, and select one of its neighboring stickers,  $k$, which is within a distance of $1.5\sigma$, as a potential partner for bond exchange, Fig. \ref{graph1}(a). We check the energy difference between the potential new bond $i$-$k$, and the old bond: $\delta U = (U_{B}(ik) + U_{LJ}(ij))-(U_{B}(ij) + U_{LJ}(ik))$. Then, we accept this potential bond swap using the Metropolis criterion. Main chain bonds are never created or destroyed in this process. Also, each sticker can participate at most in one crosslink. The monomers and stickers are only allowed to move during the MD, but not during this crosslink formation/destruction step. We equilibrate the system first by performing $5 \times 10^7$ MD steps with an integration timestep of $0.005\tau$, followed by a production cycle of \(5 \times 10^7\) MD steps ($5\times 10^{5}\tau$).  
We employ reduced units (temperatures in units of $\epsilon/k_B$, pressure in units of $\epsilon/\sigma^3$) and in these units, the reduced pressure $P = 1$  and temperatures (varied systematically between 0.6 and 2.5) are maintained by the Nos\'e-Hoover barostat and thermostat, respectively.

    First, we study how variations in the interaction strength between dissimilar moieties  can lead to different classes of phase behavior at an illustrative temperature of $T$=1. Our previous studies show that crosslinkers are spatially well-dispersed when all the non-bonding interactions are the same (``athermal") \cite{karmakar2025rouse}, as also seen in 
Figure \ref{graph1}(b), for the case where all monomers interact through the same potential. Below, we mainly focus on the WCA interaction between the stickers and monomers. These simulations reveal a distinct difference in phase behavior based on the spatial distribution of the stickers. In the annealed case we observe macrophase separation, Figure \ref{graph1}(c). The stickers aggregate forming large, continuous domains that segregate completely from the monomers. %This type of separation suggests that the thermodynamic incompatibility between stickers and monomers drives large-scale phase segregation when any bead can form crosslink, resulting in a two-phase system. 
In contrast, for quenched stickers, when the reactive bead positions are fixed along the chains — either evenly spaced, in diblock, or triblock arrangements — the system exhibits microphase separation rather than full demixing, Figure \ref{graph1}(d). %Instead of large-scale domain formation, the reactive beads self-assemble into localized clusters or pockets that remain embedded within the continuous polymer matrix. 
This difference indicates that positional control over reactive sites poses additional constraints on their ability to fully phase separate, instead favoring the formation of dispersed, self-limiting aggregates at the nanoscale. 

This structural distinction between macrophase and microphase separation is clearly illustrated by the pair correlation function of stickers for the three cases. The case of well-mixed stickers shows almost gas-like behavior, as expected, Figure \ref{graph1}(e). The WCA case involving annealed stickers shows pair correlation reminiscent of macrophase separation, while the quenched sticker cases show moderate peaks consistent with microphase separation. Further, the cluster distributions for all three cases are shown in Figure \ref{graph1}(f). Consistent with the stickers pair correlation functions, a wide range of cluster size is seen in the quenched case, whereas fewer clusters with significantly large sizes are formed for the annealed stickers.

\begin{figure}[!htb]
    \centering
     \includegraphics[width=16.3cm,height=10.3cm]{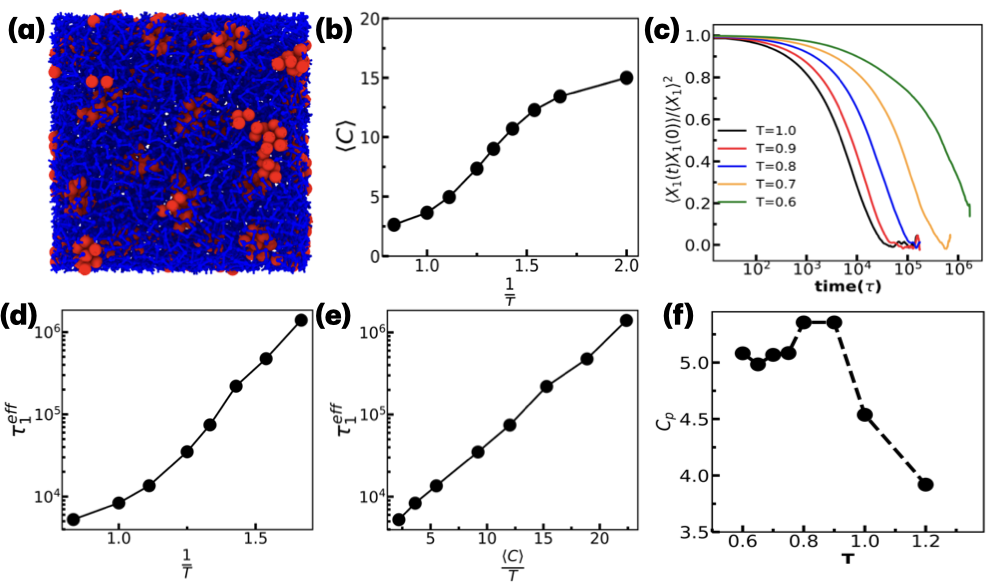}
    \caption{(a) A MD snapshot of the quenched system where stickers are evenly distributed along the polymer chains for $T=0.7$. (b) The average cluster size, $\langle C \rangle$, is plotted as a function of 1/T. (c) The end-to-end vector autocorrelation functions for several temperatures. (d) The resulting chain relaxation time is plotted as a function  of 1/T. (e) The chain relaxation time shows a log-linear dependence on $\langle C \rangle /T$. (f) The specific heat is plotted as a function of temperature. }
    \label{graph2}
\end{figure}

Here, we consider the system with quenched disorder, specifically where the stickers are equally spaced along the chain. We performed simulations across a range of temperatures focusing on two key observables: the average cluster size of stickers ($\langle C \rangle$) and the relaxation time of chains. We measure the relaxation behavior of the chains using Rouse modes ($p
= 1, 2, ..., N − 1$) of a chain of length N following $\Vec{X}_{p}=(2/N)^{1/2}\sum_{i=1}^{N} \Vec{r}_{i} cos[(p\pi/N)(i-1/2)]$\cite{kalathi2014rouse}. The modes describe internal relaxations with a mode number $p$ corresponding to a chain fragment consisting of $(N − 1)/p$ bonds, where $p=1$ corresponds to the longest mode, the end-to-end relaxation. The autocorrelation of the Rouse modes, termed as Rouse relaxations, is observed to follow a stretched exponential temporal dependence of the form $\langle \Vec{X}_{p}(t)\cdot \Vec{X}_{p}(0) \rangle = \langle \Vec{X}_{p}^{2} \rangle e^{-(\frac{t}{\tau_{p}})^{\beta_{p}}} $ in past works. Here, $\tau_{p}$ and $\beta_{p}$ are the Rouse relaxation time and the exponent of mode $p$, respectively. We compute the effective relaxation time as $\tau_{p}^{eff} = \frac{\tau_{p}}{\beta_{p}}\Gamma(1/\beta_{p})$. Here, we specifically calculate the end-to-end distance relaxation function, i.e. the mode corresponding to $p$=1 in the Rouse model.

%Here we discuss for the evenly spaced case (Case a\ref{graph3}(a)). 
An MD snapshot of the system is shown at $T=0.7$, Figure \ref{graph2}(a). From such simulations we find that the $\langle C \rangle$ displays a 
sigmoidal-like dependence on $1/T$, Figure \ref{graph2}(b), i.e., the average cluster size increases as the temperature decreases and whose rate of increase eventually saturates at lower temperatures. 
%\sscom{This description needs work: This trend reflects the thermally driven nature of aggregation — at higher temperatures, thermal energy counteracts aggregation, while at lower temperatures the WCA interactions dominate (dislike between non-reactive and reactive beads becomes profound), leading to the formation of larger and more stable clusters.} 
The initial growth of the clustering is thermodynamically driven by the unfavorable monomer-sticker interactions, but continued clustering at lower temperatures is hindered by chain connectivity. Analogous to the case of small molecule surfactancy, the need to accommodate chain connectivity restricts the maximum size that a cluster can attain.
In a similar manner, the chain relaxation time, ($\tau_{1}^{eff}$), increases as the temperature decreases, Figure \ref{graph2}c. Evidently, the activation energy for this relaxation, i.e., the local slope of a $log \tau_1^{eff}$ vs. $1/RT$, increases with decreasing temperature. We conjecture that this must reflect the increasing energy cost of "pulling" a sticker out of cluster whose size grows with decreasing temperature. To this point, while the $\tau_1^{eff}$ shows a non-linear temperature dependence on a log-linear plot, it appears to have a linear dependence on $\frac{\langle C \rangle }{T}$, Figure \ref{graph2}(e). Since the heat capacity, $C_p$ has a peak in the vicinity of $T\approx$0.85, this suggests that the aggregation transition occurs in this temperature range, Figure \ref{graph2}f. The presence of a super-Arrhenius to Arrhenius temperature dependence, together with a heat capacity maximum, is strongly reminiscent of a fragile-to-strong cross-over, which has been investigated intensely in the case of supercooled water and glass alloys  \cite{angell_ftos,wei2015phase,metglas_ftos}. A comparison of the mechanisms in the present and the earlier studied cases will be very instructive. These results strongly suggest that chain dynamics are controlled by both sticker exchange (which likely follows an Arrhenius temperature dependence) and energy cost of sticker pull out of a cluster which likely scales with the cluster size. This result, which reflects the key role of microphase separation on the dynamics of associatively crosslinked polymers is the most significant finding of this paper.

\begin{figure}[!htb]
    \centering
     \includegraphics[width=16.3cm,height=8.3cm]{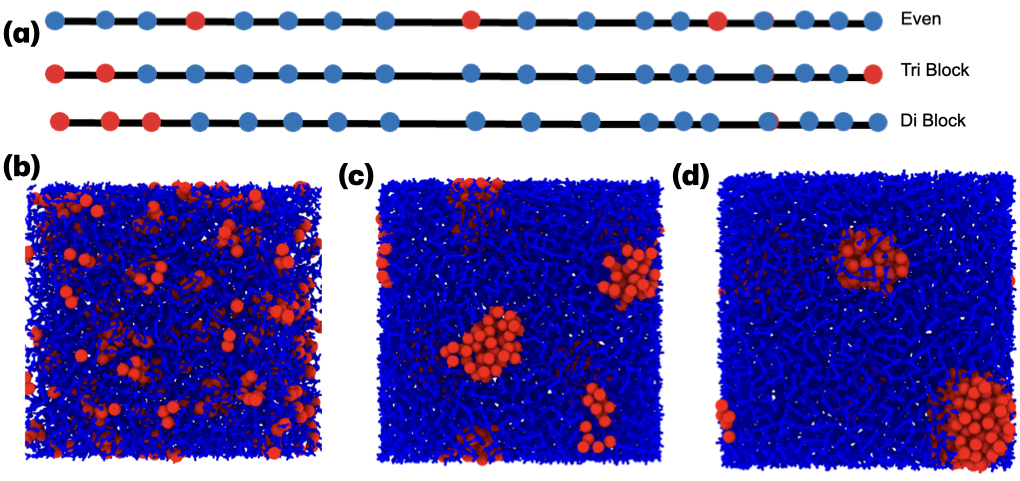}
    \caption{A schematic representation of three polymers are shown in (a). The MD snapshot of uniform, diblock and triblock copolymer systems are shown in (b), (c), and (d), respectively, for T=1.}
    \label{graph3}
\end{figure}

\begin{figure}[!htb]
    \centering
     \includegraphics[width=16.3cm,height=14.3cm]{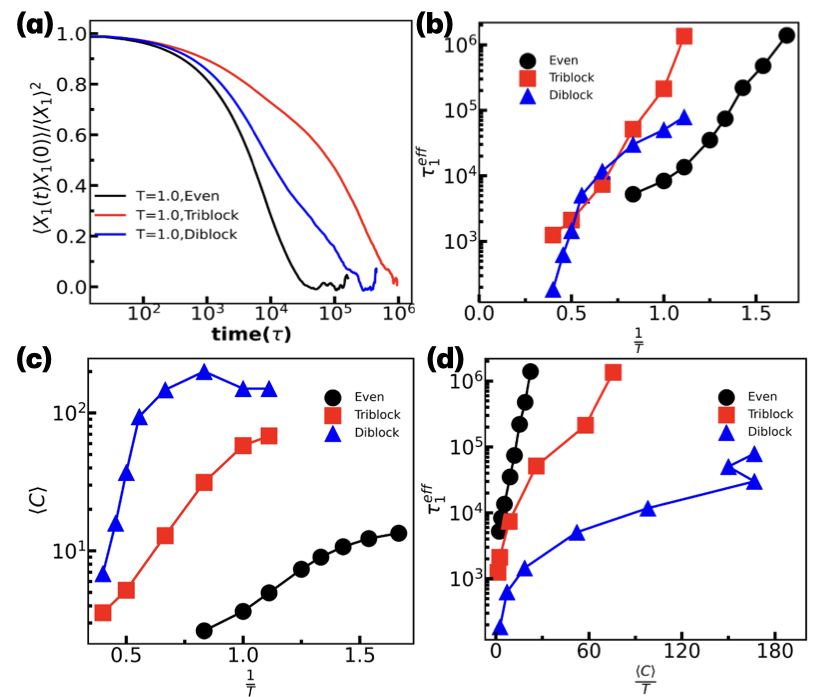}
    \caption{ The end-to-end vector autocorrelation functions are compared in (a)for uniform, diblock and triblock copolymers. The relaxation time is plotted as a function of 1/T in (b) for all the cases. The mean cluster size is shown as a function 1/T in (c). The relaxation time is plotted as a T-normalized mean cluster size in (d).}
    \label{graph4}
\end{figure}

Next, we explore the role of chain architecture in dictating the extent and nature of microphase separation, which in turn directly influences the chain relaxation dynamics. Here we only consider systems with quenched stickers. To systematically explore this effect, we studied three distinct chain architecture as shown schematicaly in Figure \ref{graph3}a: the case (a) evenly spaced stickers (already discussed), case (b) triblock, and case (c) diblock. We show respective MD snapshots for each of these cases in Figures \ref{graph3}(b-d), and compare their end-to-end relaxation behavior  at fixed temperature $T=1$, Figure \ref{graph4}(a). As expected, the case with evenly distributed stickers exhibits the fastest relaxation, while the triblock case shows the slowest — with a relaxation time difference between the two that spans about two orders of magnitude. This highlights the profound influence that the spatial arrangement of stickers has on the dynamic behavior of the network, even without altering crosslinker density or chemistry.

The relaxation time $\tau_{1}^{eff}$ evolves with inverse temperature ($1/T$) in a manner that is qualitatively similar across the three architectures, Figure \ref{graph4}(b). All three systems exhibit the expected increase in relaxation time with decreasing temperature, reflecting reduced chain mobility as clustering becomes more pronounced and the difficulty in exchanging crosslinks increases. Interestingly, the triblock case displays a temperature dependence similar to the case with evenly spaced stickers,  with the whole behavior apparently shifted to higher temperatures. This is in line with the fact that the clusters form and grow at higher $T$ for the triblocks, likely because the entropic loss on self-assembly is lower. Indeed a plot of $\tau_1^{eff}$ vs $\langle C \rangle/T$ for both architectures have a similar dependence at small values of $\langle C \rangle/T$, Figure \ref{graph4}d. The departures at larger $\langle C \rangle/T$ likely reflect the fact that the clusters in the two cases are very different in shape and size, Figures \ref{graph3}b-c, \ref{graph4}c. Thus, clearly $\langle C \rangle/T$ is not the appropriate descriptor to compare the relaxation behavior across these two architectures at low temperatures. Rather, we suspect that the energy required to pull a sticker out of a cluster saturates when a cluster becomes larger than the cutoff radius for the LJ interaction - thus, we expect that the energy for sticker pullout, rather than the cluster size is a better metric in these cases. The diblock architecture shows an even larger change in cluster size; its relaxation behavior also shows a very different character when plotted against $\langle C \rangle/T$. Since the clusters here are even larger, we conclude that using a mean cluster size alone is likely a naive means of understanding the difficulty in exchanging stickers across different clusters, a phenomenon that is critical for chain relaxation.

We investigated the structure and relaxation dynamics of a coarse-grained polymer melt with reversible crosslinks using a hybrid molecular simulation approach combining MC with MD simulations. By implementing an associative network model that conserves the total number of bonds, we explore how chain architecture and the thermodynamic compatibility between the chain monomers and stickers influence crosslinker aggregation and chain dynamics. When the stickers and monomers are compatible, then the stickers are well-mixed with the monomers. In contrast, incompatible sticker-monomer interactions lead to thermodynamic incompatibility. The spatial arrangement of stickers along polymer chains critically influences the resulting morphology. When the stickers are annealed (i.e., they can freely move on the chain) then we have macrophase separation, whereas quenched disorder (fixed stickers) induces microphase separation. The three different quenched chain architectures tested (uniform distribution of stickers, diblock, and triblock) show significant differences in crosslinker clustering and chain relaxation behavior. Temperature further tunes this behavior by controlling cluster growth and relaxation dynamics. As aggregation strengthens, chain mobility decreases, confirming that microphase separation directly affects relaxation and mechanical properties. 
Evenly spaced reactive beads facilitate faster relaxation and smaller aggregates, while asymmetric distributions promote larger clusters and slower dynamics. Notably, the Arrhenius scaling between $\tau_1^{eff}$ and $\langle C \rangle / T$ is seen only for the evenly spaced stickers case, while the diblock and triblock arrangements deviate from Arrhenius behavior as aggregation grows, indicating that the network geometry and connectivity increasingly control dynamics at larger cluster sizes. Overall, these findings show that both chemical architecture and thermal conditions dictate the degree of microphase separation and directly influence the mechanical response of dynamic covalent networks.

\section*{Acknowledgments}

This work is made possible by financial support from the SERB, DST, Government of India through a core research grant (CRG/2022/006926). This research uses the resources of the Center for Nanoscale Materials, Argonne National Laboratory, which is a DOE Office of Science User Facility supported under the Contract DE-AC02-06CH11357. SKK gratefully thanks National Science Foundation
(Division of Materials Research, DMR-2301348) for funding. SS acknowledges support through the JC Bose Fellowship (JBR/2020/000015) from the Science and Engineering Research Board, Department of Science and Technology, India.

%\bibliographystyle{vancouver}
%\bibliography{paper.bib}
\begin{comment}

\end{comment}

\end{document}